# Engineering Robust Metallic Zero-Mode States in Olympicene Graphene Nanoribbons

Ryan D. McCurdy,[†,‡] Aidan Delgado,[†,‡] Jingwei Jiang,[‡,§,‡] Junmian Zhu,[†] Ethan Chi Ho Wen,[†] Raymond E. Blackwell,[†] Gregory C. Veber,[†] Shenkai Wang,[†] Steven G. Louie,[‡,§,*] Felix R. Fischer[†,§,¶,◊,*]

[†]Department of Chemistry, University of California, Berkeley, CA 94720, U.S.A.

[‡]Department of Physics, University of California, Berkeley, CA 94720, U.S.A.

[§]Materials Sciences Division, Lawrence Berkeley National Laboratory, Berkeley, CA 94720, U.S.A.

[¶]Kavli Energy NanoSciences Institute at the University of California Berkeley and the Lawrence Berkeley National Laboratory, Berkeley, California 94720, U.S.A.

[◊]Bakar Institute of Digital Materials for the Planet, Division of Computing, Data Science, and Society, University of California, Berkeley, CA 94720, USA.

**ABSTRACT:** Metallic graphene nanoribbons (GNRs) represent a critical component in the toolbox of low-dimensional functional materials technology serving as 1D interconnects capable of both electronic and quantum information transport. The structural constraints imposed by on-surface bottom-up GNR synthesis protocols along with the limited control over orientation and sequence of asymmetric monomer building blocks during the radical step-growth polymerization has plagued the design and assembly of metallic GNRs. Here we report the regioregular synthesis of GNRs hosting robust metallic states by embedding a symmetric zero-mode superlattice along the backbone of a GNR. Tight-binding electronic structure models predict a strong nearest-neighbor electron hopping interaction between adjacent zero-mode states resulting in a dispersive metallic band. First principles DFT-LDA calculations confirm this prediction and the robust, metallic zero-mode band of olympicene GNRs (oGNRs) is experimentally corroborated by scanning tunneling spectroscopy.

## INTRODUCTION

Graphene nanoribbons (GNRs) are representatives of an emerging class of bottom-up synthesized designer quantum materials whose electronic structure can be tuned with atomic precision by deterministic chemical design. Their structures exhibit unusual and some never before realized physical properties that extend far beyond the parent 2D graphene. Highly tunable band gaps,[1-3] photoemission,[4] magnetic spin chains,[5] and even symmetry protected topological states[6-9] can all be tailored by real space structural parameters including among others width, symmetry, edge termination, and substitutional doping.[10-13] A dominant electronic feature common to almost all GNRs is the opening of a sizeable band gap imposed by laterally confining a 2D graphene sheets to a quasi-1D GNR (width < 2 nm). This quantum confinement effect has emerged as a veritable challenge to the design of intrinsically metallic band structures. Bottom-up access to a family of robust metallic GNRs not only represents a critical component in the development of advanced nanographene based logic circuits,[14] e.g. as covalent interconnects capable of electronic and quantum transport, but could serve a versatile and highly tunable platform to explore emergent physical phenomena such as Luttinger liquids,[15-18] plasmonics,[19-22] charge density waves,[23-26] and superconductivity in 1D.[27-30]

We recently reported a general approach for accessing metallic GNRs by embedding a superlattice of localized zero-mode states along the backbone of a bottom-up synthesized sawtooth GNR (sGNR).[31-32] A key ingredient to this approach was the design of a molecular building block, 6,11-bis(10-bromoanthracen-9-yl)-1-methyltetracene (BAMT in Figure 1), that introduces a sublattice imbalance ($\Delta N = N_A - N_B$) between carbon atoms occupying the A and the B sublattice sites of graphene, respectively. The concept is reminiscent of Lieb's theorem,[33] a surplus of carbon atoms on sublattice A versus sublattice B will lead to $\Delta N$ eigenstates at $E = 0$ eV, or zero-modes, localized on the majority sublattice. Application of a simple tight binding model, the Su-Schrieffer-Heeger (SSH) dispersion relationship,[34] that describe the interaction between these local zero-mode states gave rise to two distinctive bands defined by an intracell hopping amplitude $t_1$ and an intercell hopping amplitude $t_2$. The energy gap enclosed by these bands is $\Delta E = 2||t_1|-|t_2||$. If the absolute magnitude of the two hopping amplitudes are equal, i.e. $|t_1| = |t_2|$, as illustrated for the evenly spaced zero-mode states in sGNR (Figure 1A) the energy gap vanishes and the 1D electronic structure becomes metallic.[35-36] The presence of a metallic zero-mode band at the Fermi level ($E_F$) in sGNRs could be visualized by scanning tunneling spectroscopy (STS) and was further corroborated by DFT-LDA calculations. This method, however, suffered from a Stoner-type instability for narrow bands that could open up a spin-splitting gap. To overcome this, we had to introduce an effective sublattice mixing (e.g. introduction of 5-membered-rings in 5-sGNRs) to facilitate the hopping between the localized zero modes.

A major shortcoming inherent to the design of 5-sGNRs is the requirement that all bonds formed between molecular precursors as part of the on-surface radical step-growth polymerization have to follow a strict head-to-tail pattern (–AB–AB–AB– in Figure 1A) to ensure the intracell hopping amplitude $|t_1|$ remains equal in magnitude to the intercell hopping amplitude $|t_2|$. The statistical probability that this specific arrangement is adopted for a single C–C bond forming step on the surface is only ~50%. Were the molecular building blocks to fuse in the undesirable head-to-head (–BA–AB–) or tail-to-tail (–AB–BA–) configuration the zero-mode bands would split ($|t_3| \neq |t_4|$) and give rise to a semiconductor rather than a metal.[31-32] The probability of producing a metallic sGNR segment from $n$ monomers is therefore $P_n = (0.5)^n$ or less than 1% for $n > 7$, severely limiting the use of metallic sGNRs at length scales necessary for applications as device interconnects. While sGNRs served as a successful proof-of-concept for our general approach to access metallic phases in GNRs, designs that ensure regioregularity and an efficient sublattice mixing of zero-mode states are needed to obtain uniform samples of extended GNRs with persistent, intrinsically metallic zero-mode bands.

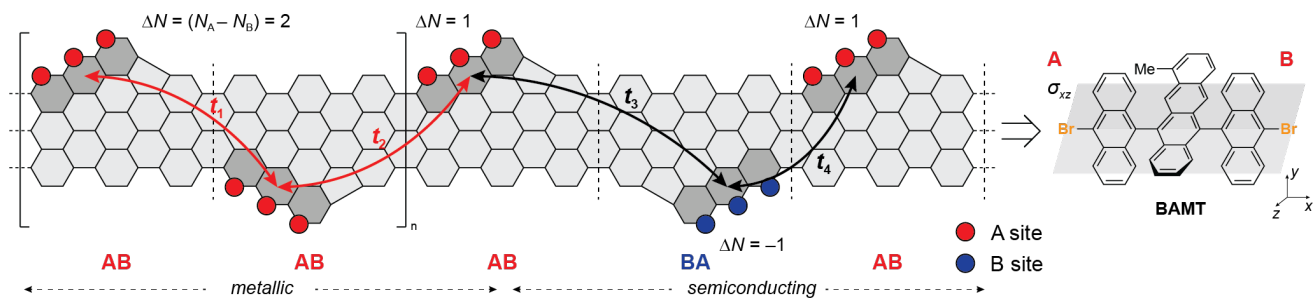

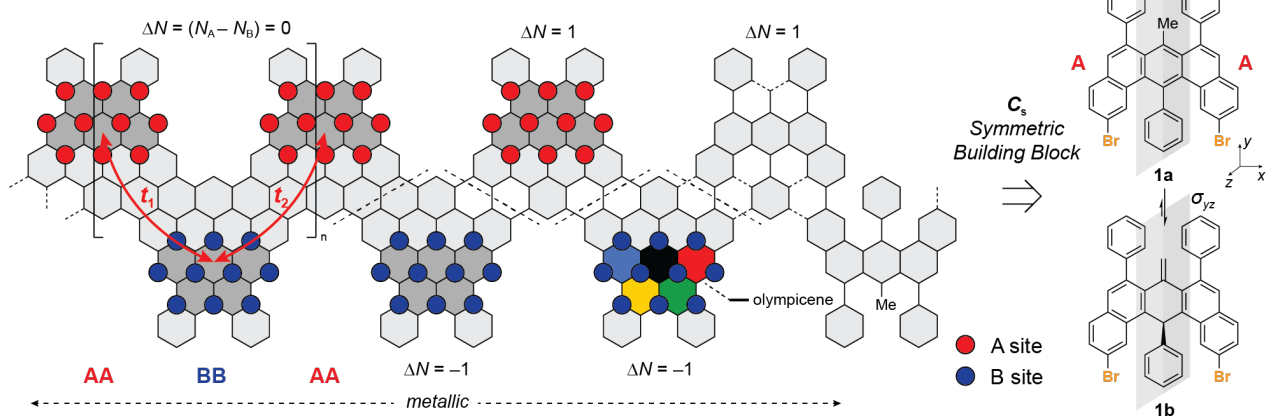

**Figure 1.** Bottom-up design and zero-mode engineering of metallic GNRs. (A) The metallic band in 5-sGNRs emerges only if the orientation of the monomers along the axis of polymerization ($x$-axis) follows a regioregular –AB–AB–AB– pattern. (B) The presence of a $\sigma_{yz}$ mirror plane in **1a/b** normal to the axis of polymerization ($x$-axis) ensures that either of two possible orientations of a monomer during the radical step-growth polymerization gives rise to a metallic zero-mode band in oGNRs.

Here we report the design and on-surface synthesis of metallic olympicene GNRs (oGNRs) derived from the $C_s$ symmetric molecular building block **1b** (Figure 1B). Rather than relying on a statistical distribution of bond forming events that dictated the band structure in sGNRs, the molecular building blocks for oGNRs feature a $\sigma_{yz}$ mirror plane perpendicular to the $x$-axis, the main axis of polymerization, ensuring that oGNRs arising from **1b** will always be metallic. This could be achieved by placing the carbon atom contributing to the sublattice imbalance $\Delta N$, the methyl group in **1a** or the methylene in **1b**, along the $\sigma_{yz}$ mirror plane of the building block. The arrangement of any two monomers forming the oGNR unit cell ensures that the position of the zero-mode state alternates between the A and the B sublattice sites. The efficient sublattice mixing that gives rise to a robust metallic zero-mode band is built into the design. Atomically precise oGNR were synthesized from molecular precursors on a Au(111) surface and characterized in ultrahigh-vacuum (UHV) by low-temperature scanning tunnelling microscopy (STM) and spectroscopy (STS). Experimental results are further corroborated by first-principles calculations revealing a robust metallic band that spans across $E_F$ emerging from the interaction of zero-mode states along the backbone of oGNRs.

## RESULTS AND DISCUSSION

### Synthesis of Molecular Precursors for oGNRs.
The synthesis of the molecular precursor **1b** for oGNRs is depicted in Figure 2. Double Suzuki cross-coupling of 2,6-dibromo-4-methyl-1,1'-biphenyl (**2**) with two equivalents of 2-(5-methoxy-2-(phenylethynyl)phenyl)-4,4,5,5-tetramethyl-1,3,2-dioxaborolane (**3**) yielded the diyne **4**. Treatment of **4** with Barluenga's reagent in TfOH successfully induced the sterically demanding benzannualtion to give the benzo[$m$]tetraphene core **5**. The two aryl iodide groups in **5** were removed by lithium-halogen exchange with $s$-BuLi followed by protonation with MeOH to yield **6**. With the assembly of the characteristic carbon backbone of the monomer building block completed, the task shifted to converting the methoxy groups in **6** to aryl halides that serve as thermally labile chemical handles during the on-surface GNR growth. A well-precedented route involves deprotection of aryl-methyl ethers to reveal the free alcohols followed by conversion into aryltriflates which serve as versatile handles for further diversification.[11] [1]H and [13]C NMR revealed that deprotection of **6** under Lewis/Brønsted acidic (e.g. BBr$_3$, AlBr$_3$, TMSI, HBr, HI, TfOH) or nucleophilic (e.g. NaSEt, LiI) conditions induced a tautomerization of the benzo[$m$]tetraphene core to yield exclusively the 7-

methylene-7,14-dihydrobenzo[*m*]tetraphene **7b** rather than the anticipated tautomeric species **7a**. Following the synthetic route outlined above treatment of **7b** with Tf₂O gave access to the triflate **8b**. Single crystals suitable for X-ray diffraction were grown by slow diffusion of MeOH into a saturated solution of **8b** in CH₂Cl₂. The crystal structure of **8b** revealed that the central ring of the dihydrobenzo[*m*]tetraphene core, ring **c** in Figure 2, adopts a boat-like conformation placing the methylene group at C7 and the phenyl group at C14 at an angle of 35.0° and 76.0° above the base plane spanned by the remaining four carbon atoms (C6a, C7a, C13b, C14a) of ring **c**, respectively. While this conformation comes at the cost of breaking the extended aromatic ring-system of a benzo[*m*]tetraphene core into two isolated naphthalene units, the boat conformation adopted by ring **c** significantly reduces the A$^{1,3}$ strain between the exocyclic methylene group and the two phenyl substituents at C6 and C8. To complete the synthesis the triflates in **8b** were converted into the diboronic ester **9b** before treatment with excess CuBr₂ yielded the 2,12-dibromo-7,14-dihydrobenzo[*m*]tetraphene **1b**, the molecular building block for oGNRs. Single crystals of **1b** suitable for X-ray diffraction and surface-assisted oGNR growth were obtained by diffusion of MeOH into a saturated solution of **1b** in CH₂Cl₂. In close analogy to the conformation adopted by **8b**, the ring **c** in dihydrobenzo[*m*]tetraphene **1b** adopts a boat-like conformation. The included angles between the methylene group at C7 and the phenyl substituent at C14 with the base plane of ring **c** are 37.5° and 75.0°, respectively.

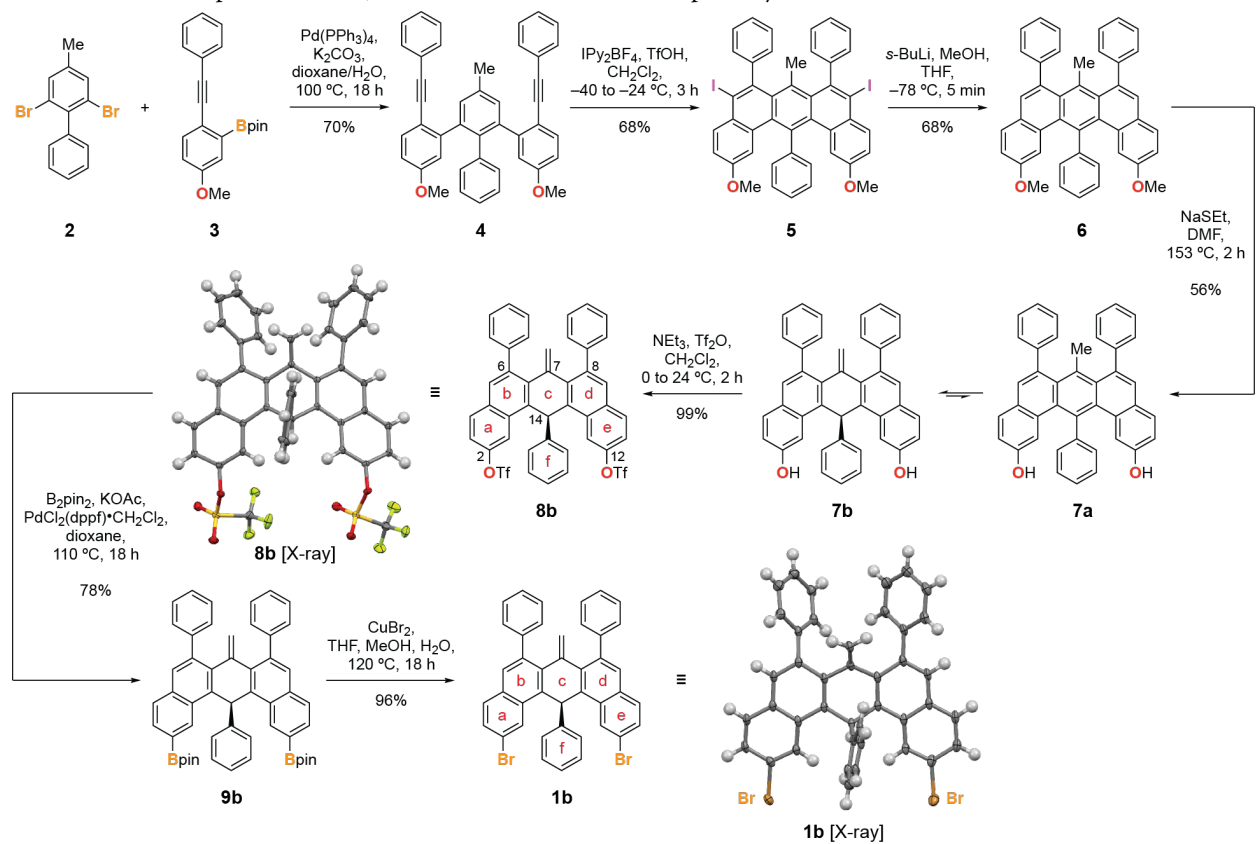

**Figure 2.** Synthesis of molecular precursor **1b** for oGNRs. Single X-ray crystal structures of **8b** and **1b**. Thermal ellipsoids are drawn at the 50% probability level. Color coding: C (gray), O (red), F (green), S (yellow), Br (orange). Hydrogen atoms are placed at calculated positions.

**Surface Assisted Growth and Electronic Structure Characterization of oGNRs.** Samples of metallic oGNRs were prepared following an established surface-assisted bottom-up GNR synthesis. Molecular precursor **1b** was sublimed in UHV from a Knudsen cell evaporator onto a Au(111) surface held at 25 °C. Figure 3A shows a representative topographic STM image of self-assembled islands of **1b** on an atomically flat Au(111) terrace. Step growth polymerization of **1b** was induced by annealing the molecule-decorated surface first to 180 °C for 15 min followed by a second annealing step at 350 °C for 15 min to complete the cyclodehydrogenation. Topographic images of a high and low coverage sample, Figure 3B and 3C, respectively, reveal extended GNRs featuring a characteristic alternating pattern of protrusions along the backbone of the GNR and lengths ranging up to 30 nm (Supporting Information Figure S1). Bond-resolved STM (BRSTM) with CO-functionalized tips reveals that following the initial radial step growth polymerization at 180 °C the [4]-helicene fragments lining the edges of oGNRs have partially fused to form 5-membered rings (Figure 3E). The second annealing step (350 °C for 15 min) merely completes the process giving access to a uniform edge termination in 5-oGNRs (Figure 3D, Supporting Information Figure S2).

Having resolved the chemical structure of 5-oGNRs we shifted our focus to the characterization of its local electronic structure using differential tunneling spectroscopy. Figure 4A shows typical d$I$/d$V$ point spectra for a 5-oGNR recorded with a CO-functionalized STM tip at the positions highlighted in the inset. Three spectral features can clearly be seen in the range of –2.00 V < $V_s$ < +1.80 V. Two shoulders at $V_s$ = +1.60 V (*Peak* 1) and $V_s$ = –0.75 V (*Peak* 3) dominate the spectrum, along with a broad peak centered at $V_s$ = –0.90 V (*Peak* 2). The signal intensities of *Peaks* 1 and 3 are strongest when the STM tip is placed close to the convex protrusions lining the edge of the ribbon (blue line in Figure 4A), whereas *Peak* 2 is prominently featured in both spectra recorded above the center of an olympicene unit (red line in Figure 4A, Supporting Information Figure S3) and along the edge of the ribbon. Figure 4B shows a magnification of the d$I$/d$V$ spectra taken over a narrower bias range –0.20 V < $V_s$ < +0.20 V. Most prominent here is a U-shaped feature anchored by two peaks in the differential conductance spectrum at $V_s$ = –0.10 V and $V_s$ = +0.10 V when the STM tip is placed above the center of the ribbon. Differential conductance maps recorded over a continuous bias range of $V_s$ = +0.10 V to $V_s$ = –0.10 V (Figure 4D–J) show that the same state, associated with a distinctive wavefunction pattern of a zero-mode, spans across $E_F$. The peak at $V_s$ = –0.10 V can thus be assigned to the bottom edge of the lower (LZM) of two zero-mode (ZM) bands contributing to the metallic state in 5-oGNRs, while the peak at $V_s$ = +0.10 V captures the top edge of the upper zero-mode (UZM) band. The U-shaped LDOS spanning across $E_F$ is the signature of van Hove singularities associated with the flat band-edges of the LZM and UZM bands.

First-Principles Calculation of 5-oGNR Electronic Structure. We further explored the metallic band structure of 5-oGNRs using ab initio density functional theory (DFT). Figure 4M and 4N show the theoretical DOS and the band structure of a 5-oGNR calculated using a local density approximation (LDA) to the exchange-correlation potential. Two highly dispersive bands, labeled LZM and UZM, span across the energy scale from $E–E_F$ = –0.25 V to $E–E_F$ = +0.25 V. The LZM and UZM bands cross $E_F$ at $\boldsymbol{k}$ = $X$ giving rise to a robust metallic band with a width of $E_{ZM}$ ~ 0.5 eV. Both the upper and lower edge of the ZM bands show a flattening as they approach $\boldsymbol{k}$ = $\Gamma$. The corresponding calculated DOS (Figure 4M) faithfully reproduces the U-shaped signature of the metallic band identified in the experimental LDOS (Figure 4A,B). DFT-LDA LDOS maps evaluated at the energy position of the UZM and LZM edges (Figure 4P,Q) show the characteristic nodal pattern observed in the corresponding d$I$/d$V$ maps (Figure 4D–J). At energies above and below $E–E_F$ = ±0.25 V the calculated metallic ZMB gives way to minigaps, narrow regions of vanishing DOS that span the energy window separating the ZMB from the bottom of the CB and the top of the VB, located at $E–E_F$ = +0.80 V and $E–E_F$ = –0.75 V, respectively. Both LZM and UZM bands can be fit to an SSH tight binding model

$$E_{\pm}(k) = \pm\sqrt{|t_1|^2 + |t_2|^2 + 2|t_1||t_2|\cos(k + \delta)} \quad \text{(eq. 1)}$$

with the intra- and intercell hopping amplitudes $|t_1| = |t_2| = 111$ meV, and $\delta = 0$ ($\delta$ is the relative phase between $t_1$ and $t_2$). Supercell calculations further show that the rigid GNR backbone renders oGNR virtually impervious to mechanical deformations usually associated with strong electron-phonon coupling along the main $x$-axis of the ribbon that would otherwise induce spontaneous metal-insulator transitions (i.e. Peierls distortion). Besides the decisive structural advantage over the first generation metallic sGNRs, the $C_s$ symmetric molecular precursor **1b** features a $\sigma_{yz}$ mirror plane perpendicular to the axis of polymerization that gives rise to uniform predictable monomer sequences that exclusively yield metallic band structures, the family of oGNRs has one last trick up its sleeve.

Unlike sGNRs, where a fusion of the [4]helicene fragment along the sawtooth edge proved critical to induce a mixing of the sublattice spin polarized zero-mode states that led to a broadening of the metallic ZMB (i.e. a reduced DOS at $E_F$) sufficient to circumvent Mott insulator or Stoner magnetic phase transitions, an efficient hopping between zero-mode states localized on A and B sites in oGNRs is built into our design. The –AA–BB–AA– polymerization places zero-mode states on alternating sublattice sites ensuring that the hopping amplitudes $t_1$ and $t_2$ between adjacent states is dominated by the nearest neighbor hopping term rather than the much smaller second nearest neighbor hopping (Figure 1B). This is reflected in band structure calculations using the local spin density approximation (LSDA) that show no sign of magnetic phase transitions for the disperse metallic ZM bands in 5-oGNRs (Supporting Information Figure S4).

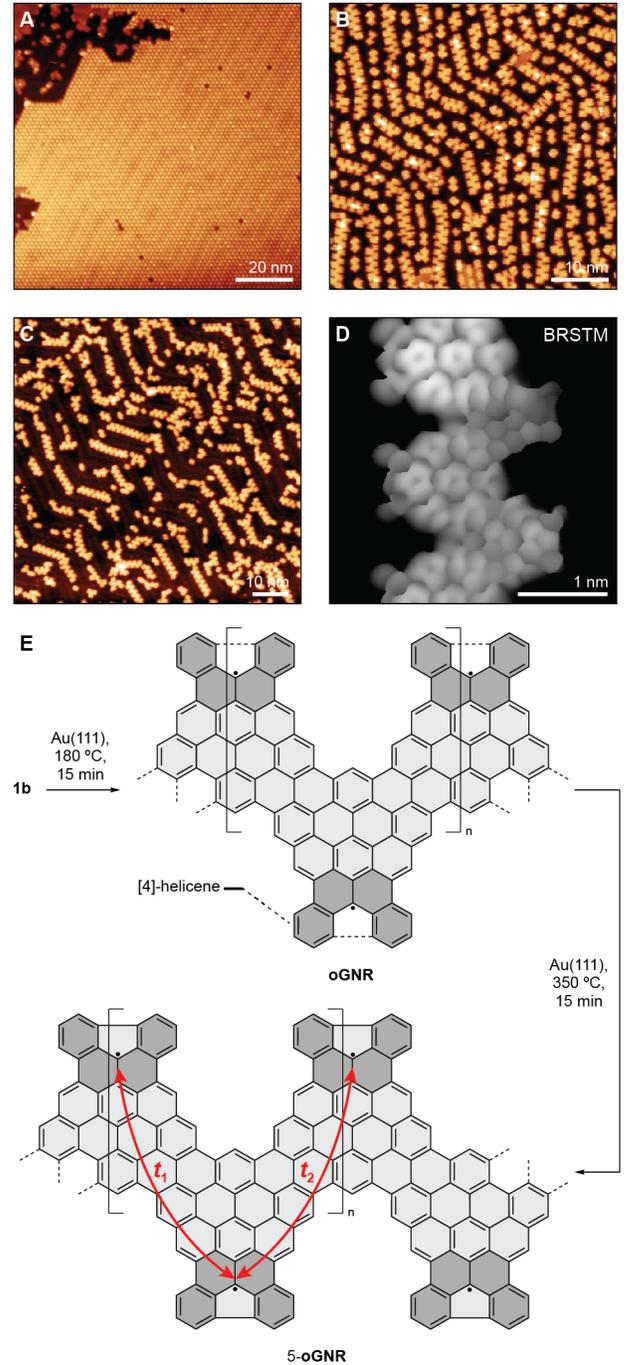

Figure 3. Bottom-up synthesis of 5-oGNRs. (A) STM topographic image of a self-assembled island of molecular precursor **1b** on Au(111) ($V_s$ = 0.05 V, $I_t$ = 20 pA). (B) STM topographic image of a high coverage sample of 5-oGNRs following annealing to 350 °C ($V_s$ = 0.05 V, $I_t$ = 20 pA). (C) STM topographic image of a low coverage sample of 5-oGNRs following annealing to 350 °C ($V_s$ = 0.05 V, $I_t$ = 20 pA). (D) BRSTM image of a 5-oGNR segment showing the 5-membered rings resulting from the fusion of [4]helicene groups along the oGNRs edges ($V_s$ = 0.01 V, $I_t$ = 400 pA). (E) Schematic representation of the stepwise thermally induced cyclodehydrogenation that gives rise to 5-oGNRs.

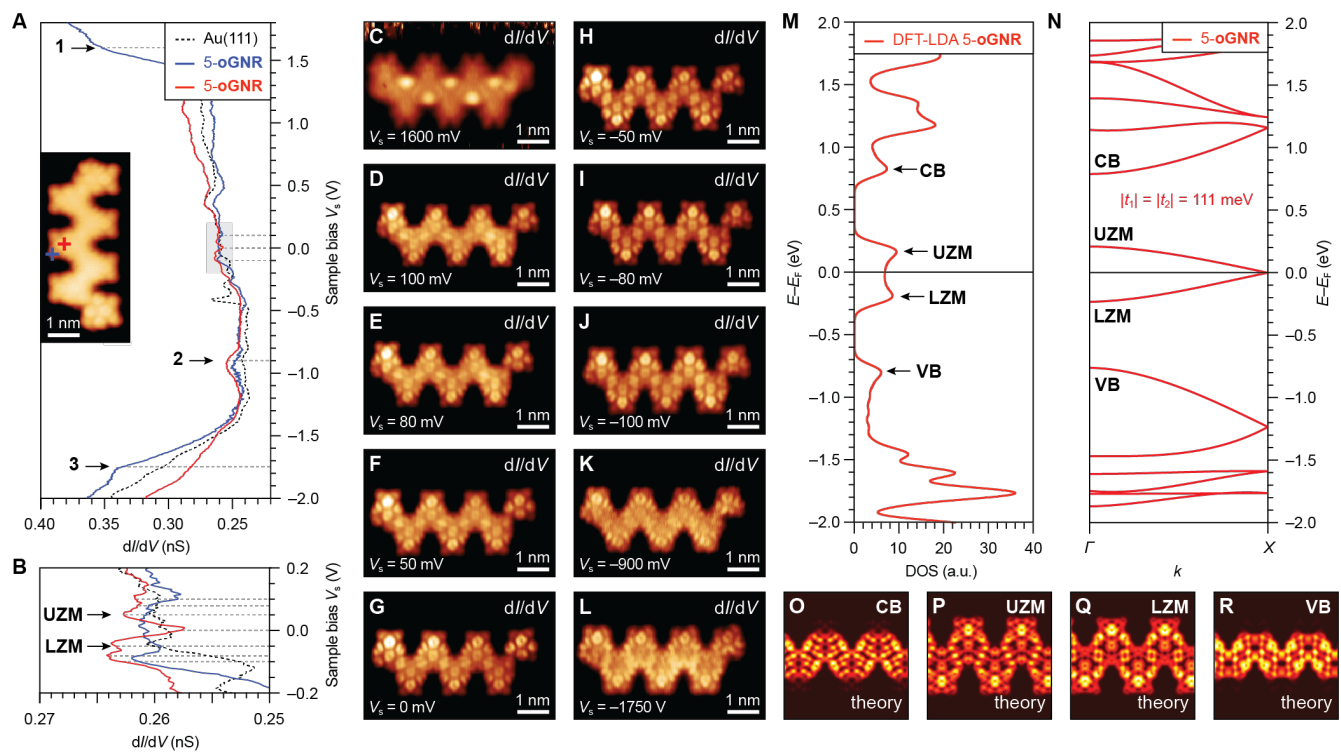

**Figure 4.** Electronic structure of 5-oGNRs. (A–B) STS d$I$/d$V$ spectra recorded on a 5-oGNR at the positions marked in the inset STM topographic image with a red and blue cross (spectroscopy: $V_{ac}$ = 11 mV, $f$ = 455 Hz; imaging: $V_s$ = 50 mV, $I_t$ = 20 pA, CO-functionalized tip). (C–L) Constant height d$I$/d$V$ maps recorded at the indicated biases (spectroscopy: $V_{ac}$ = 11 mV, $f$ = 455 Hz). (M) DFT-LDA calculated DOS of 5-oGNR (spectrum broadened by 10 meV Gaussian). Features associated with the CB, UZM, LZM, and VB are indicted by arrows. (N) DFT-LDA calculated band structure of a freestanding 5-oGNRs. A tight binding fit to DFT-LDA band structure yields the hopping parameters $|t_1| = |t_2|$ = 111 meV. (O–R) Calculated DFT-LDA LDOS maps evaluated at the edge of the bulk conduction band, at the UZM and LZM bands, and at the edge of the bulk valence band.

## CONCLUSIONS

We herein demonstrate the versatility of zero-mode engineering for introducing robust metallicity in 1D GNRs. A $C_s$ symmetric molecular building block undergoes a regiocontrolled on-surface polymerization to yield homogenous samples of 5-oGNRs featuring a symmetric superlattice of zero-mode states along the GNR backbone. Guided by elementary tight-binding analysis we pioneer the design of 5-oGNRs around a strong nearest-neighbor hopping interaction between electrons in adjacent zero-mode states giving rise to a large zero-mode bandwidth that is insensitive to Peierls and Stoner metal-insulator transitions. First-principles DFT-LDA calculations and scanning tunneling spectroscopy corroborate the emergence of metallic zero-mode bands in 5-oGNRs. The design and synthesis of robust, metallic GNRs paves the way towards the realization of energy efficient integrated circuit architectures based on low-dimensional carbon materials that are capable of high-speed electronic[37-38] and quantum information processing.[39-40]

## EXPERIMENTAL SECTION

**Materials and Instrumentation.** Unless otherwise stated, all manipulations of air and/or moisture sensitive compounds were carried out in oven-dried glassware, under an atmosphere of $N_2$. All solvents and reagents were purchased from Alfa Aesar, Spectrum Chemicals, Acros Organics, TCI America, and Sigma-Aldrich and were used as received unless otherwise noted. Organic solvents were dried by passing through a column of alumina and were degassed by vigorous bubbling of $N_2$ through the solvent for 20 min. Flash column chromatography was performed on SiliCycle silica gel (particle size 40–63 μm). Thin layer chromatography was carried out using SiliCycle silica gel 60 Å F-254 precoated plates (0.25 mm thick) and visualized by UV absorption. All $^1$H and $^{13}$C{$^1$H} NMR spectra were recorded on a Bruker AV-600 spectrometer, and are referenced to residual solvent peaks ($CD_2Cl_2$ $^1$H NMR = 5.32 ppm, $^{13}$C{$^1$H} NMR = 53.84 ppm). ESI mass spectrometry was performed on a Finnigan LTQFT (Thermo) spectrometer in positive ionization mode. X-ray crystallography was performed on a Rigaku XtaLAB P200 equipped with a MicroMax 007HF dual-source rotating anode and a Pilatus 200K hybrid pixel array detector. Data were collected using Mo-Kα ($\lambda$ = 0.71073 Å) radiation. Crystals were kept at 100 K throughout the collection using an Oxford Cryostream 700 for **1b** and **8b**. Data collection was performed with CrysAlisPro.[41] Data was processed with CrysAlisPro and includes a multi-scan absorption correction applied using the SCALE3 ABSPACK scaling algorithm within CrysAlisPro. Crystallographic data was solved with ShelXT, refined with ShelXL and finalized in Olex1.5.

**2-(5-methoxy-2-(phenylethynyl)phenyl)-4,4,5,5-tetramethyl-1,3,2-dioxaborolane (2).** A 50 mL Schlenk flask was charged under $N_2$ with 2-bromo-4-methoxy-1-(phenylethynyl)benzene (0.500 g, 1.75 mmol), bis(pinacolato)diboron (0.670 g, 2.63 mmol), and potassium acetate (0.515 g, 5.25 mmol) in dry dioxane (10 mL). The reaction mixture was degassed by sparging with $N_2$ for 20 min before [1,1'-bis(diphenylphosphino)ferrocene]dichloropalladium(II) (0.07 g, 0.09 mmol) was added under $N_2$. A reflux condenser was attached and the reaction mixture stirred under $N_2$ for 18 h at 80 °C. The reaction mixture was concentrated on a rotary evaporator. Column chromatography ($SiO_2$; $CH_2Cl_2$) yielded **2** (0.570 g, 1.7 mmol, 97 %) as a colorless solid. $^1$H NMR (600 MHz, $CD_2Cl_2$) $\delta$ = 7.55 (d, $J$ = 8.0 Hz, 2H), 7.49 (d, $J$ = 8.0 Hz, 1H), 7.38–7.33 (m, 3H), 7.29 (d, $J$ = 2.0 Hz, 1H), 6.97 (dd, $J$ = 8.0 Hz, $J$ = 2.0 Hz, 1H), 3.85 (s, 3H), 1.39 (s, 12H) ppm; $^{13}$C{$^1$H} NMR (151 MHz, $CD_2Cl_2$) $\delta$ = 159.5, 134.4, 131.8, 128.9, 128.3, 124.8, 120.6, 117.1, 91.1, 90.0, 84.6, 83.8, 55.9, 25.3 ppm; HRMS (ESI-TOF) $m/z$: [$C_{21}H_{24}O_3B_1$]$^+$ calcd. [$C_{21}H_{24}O_3B_1$] 335.1813; found 335.1815.

**2,6-dibromo-4-methyl-1,1'-biphenyl (3).** A 250 mL Schlenk flask was charged under $N_2$ with $N,N$-diisopropylethylamine (2.0 g, 20 mmol) in dry THF (140 mL). The reaction mixture was cooled to −78 °C and stirred for 20 min. n-BuLi (6.2 mL, 15.5 mmol, 2.5 M in hexanes) was added dropwise and stirred for 5 min. 3,5-dibromotoluene (3.75 g, 15 mmol) was added dropwise and the reaction stirred for 20 min. $ZnCl_2$ (2.10 g, 15.5 mmol) was added and the reaction mixture stirred for 2.5 h at 24 °C. Iodobenzene (1.00 g, 5 mmol) and tetrakis(triphenylphosphine)palladium(0) (0.82 g, 0.71 mmol) were added and the reaction mixture was stirred for 18 h at 24 °C. The reaction mixture was concentrated on a rotary evaporator, diluted with H$_2$O (200 mL), and extracted with CH$_2$Cl$_2$ (300 mL). The combined organic phases were washed with H$_2$O (100 mL), saturated aqueous NaCl (100 mL), dried over MgSO$_4$, and concentrated on a rotary evaporator. Column chromatography (SiO$_2$; hexane) yielded **3** (1.60 g, 4.9 mmol, 98 %) as a colorless crystalline solid. $^1$H NMR (600 MHz, CD$_2$Cl$_2$) $\delta$ = 7.49 (s, 2H), 7.47–7.41 (m, 3H), 7.20 (d, $J$ = 8.0 Hz, 2H), 2.36 (s, 3H) ppm; $^{13}$C{$^1$H} NMR (151 MHz, CD$_2$Cl$_2$) $\delta$ = 141.7, 141.3, 140.4, 133.0, 130.0, 128.7, 128.5, 124.4, 20.8 ppm; HRMS (EI-TOF) $m/z$: [C$_{13}$H$_{10}$Br$_2$]$^+$ calcd. [C$_{13}$H$_{10}$Br$_2$] 325.9129; found 325.9125.

**5-methoxy-3'-(5-methoxy-2-(phenylethynyl)phenyl)-5'-methyl-2-(phenylethynyl)-1,1':2',1''-terphenyl (4).** A 1000 mL Schlenk flask was charged with **2** (1.45 g, 4.3 mmol), **3** (6.48 g, 19.4 mmol), and K$_2$CO$_3$ (3.57 g, 25.8 mmol) in dioxane (250 mL) and H$_2$O (40 mL). The reaction mixture was degassed by sparging with N$_2$ for 20 min before tetrakis(triphenylphosphine)palladium(0) (0.50 g, 0.43 mmol) was added under N$_2$. A reflux condenser was attached and the reaction mixture stirred under N$_2$ for 18 h at 100 °C. The reaction mixture was concentrated on a rotary evaporator, diluted with H$_2$O (200 mL), and extracted with CH$_2$Cl$_2$ (300 mL). The combined organic phases were washed with H$_2$O (100 mL), saturated aqueous NaCl (100 mL), dried over MgSO$_4$, and concentrated on a rotary evaporator. Column chromatography (SiO$_2$; 3:2 hexane/CH$_2$Cl$_2$) yielded **4** (1.75 g, 3.0 mmol, 70%) as a light-yellow solid. $^1$H NMR (600 MHz, CD$_2$Cl$_2$) $\delta$ = 7.45 (s, 2H), 7.35 (d, $J$ = 8.0 Hz, 2H), 7.30–7.06 (m, 10H), 6.97–6.86 (m, 5H), 6.70 (d, $J$ = 8.0 Hz, 2H), 6.59 (m, 2H), 3.58 (s, 6H), 2.53 (s, 3H) ppm; $^{13}$C{$^1$H} NMR (151 MHz, CD$_2$Cl$_2$) $\delta$ = 159.4, 146.7, 140.7, 140.0, 138.1, 136.2, 133.5, 131.6, 128.8, 128.2, 127.2, 126.3, 124.4, 116.6, 115.7, 113.7, 91.7, 90.1, 55.8, 21.5 ppm; HRMS (ESI-TOF) $m/z$: [C$_{43}$H$_{33}$O$_2$]$^+$ calcd. [C$_{43}$H$_{33}$O$_2$] 581.2475; found 581.2477.

**5,9-diiodo-2,12-dimethoxy-7-methyl-6,8,14-triphenylbenzo[$m$]tetraphene (5).** A 500 mL Schlenk flask was charged in the dark under N$_2$ with bis(pyridine)iodonium tetrafluoroborate (2.25 g, 6.0 mmol) in dry CH$_2$Cl$_2$ (240 mL). Trifluoromethane sulfonic acid was added dropwise and the reaction mixture stirred for 15 min at 24 °C. The reaction mixture was cooled to –40 °C before **4** (1.00 g, 1.7 mmol) was added as a solution in CH$_2$Cl$_2$ (60 mL). The reaction mixture was stirred for 30 min at –40 °C before being warmed to 24 °C over 1.5 h. The reaction mixture was diluted with saturated aqueous Na$_2$S$_2$O$_3$ (200 mL), and extracted with CH$_2$Cl$_2$ (300 mL). The combined organic phases were washed with H$_2$O (100 mL), saturated aqueous NaCl (100 mL), dried over MgSO$_4$, and concentrated on a rotary evaporator. The crude solid was dissolved in a minimum amount of CH$_2$Cl$_2$, filtered over a short pad of SiO$_2$, and precipitated by trituration with MeOH yielding **5** (0.971 g, 1.17 mmol, 68%) as a yellow solid. $^1$H NMR (600 MHz, CD$_2$Cl$_2$) $\delta$ = 8.20 (d, $J$ = 8.0 Hz, 2H), 7.55 (d, $J$ = 8.0 Hz, 2H), 7.51–7.47 (m, 3H), 7.30–7.29 (m, 6H), 7.14–7.13 (m, 4H), 6.96 (dd, $J$ = 8.0 Hz, $J$ = 2.0 Hz, 2H), 6.77 (d, $J$ = 2.0 Hz, 2H), 3.24 (s, 6H), 1.39 (s, 3H) ppm; $^{13}$C{$^1$H} NMR (151 MHz, CD$_2$Cl$_2$) $\delta$ = 157.1, 148.8, 144.9, 141.7, 135.2, 134.1, 133.5, 133.4, 132.4, 131.8, 131.4, 130.7, 130.0, 129.1, 128.4, 128.3, 127.9, 117.7, 111.7, 108.5, 55.5, 23.2 ppm; HRMS (ESI-TOF) $m/z$: [C$_{43}$H$_{30}$O$_2$I$_2$]$^+$ calcd. [C$_{43}$H$_{30}$O$_2$I$_2$] 832.0330; found 832.0331.

**2,12-dimethoxy-7-methyl-6,8,14-triphenylbenzo[$m$]tetraphene (6).** A 500 mL Schlenk flask was charged under N$_2$ with **5** (0.95 g, 1.14 mmol) in dry THF (120 mL). The reaction mixture was cooled to –78 °C and stirred for 20 min. *s*-BuLi (16.3 mL, 22.8 mmol, 1.4 M in cyclohexane) was added dropwise and the reaction mixture was stirred for 5 min at –78 °C. The reaction mixture was quenched by rapid addition of MeOH (10 mL). The reaction mixture was concentrated on a rotary evaporator, diluted with H$_2$O (200 mL), and extracted with CH$_2$Cl$_2$ (300 mL). The combined organic phases were washed with H$_2$O (100 mL), saturated aqueous NaCl (100 mL), dried over MgSO$_4$, and concentrated on a rotary evaporator. The crude solid was dissolved in a minimum amount of CH$_2$Cl$_2$, filtered over a short pad of SiO$_2$, and concentrated on a rotary evaporator. The crude solid was sonicated in a minimum amount pentane, filtered, and washed with a minimum amount of pentane yielding **6** (0.450 g, 0.77 mmol, 68%) as a yellow solid. $^1$H NMR (600 MHz, CD$_2$Cl$_2$) $\delta$ = 7.67 (d, $J$ = 8.0 Hz, 2H), 7.60 (m, 2H), 7.55 (m, 2H), 7.50 (m, 3H), 7.31 (m, 8H), 7.26 (m, 2H), 6.96 (dd, $J$ = 8.0 Hz, $J$ = 2.0 Hz, 2H), 6.81 (d, $J$ = 2.0 Hz, 2H), 3.25 (s, 6H), 1.78 (s, 3H) ppm; $^{13}$C{$^1$H} NMR (151 MHz, CD$_2$Cl$_2$) $\delta$ = 156.5, 146.0, 145.8, 136.3, 134.7, 134.0, 132.7, 131.8, 131.4, 130.8, 130.5, 129.5, 129.1, 129.1, 128.8, 128.4, 127.7, 127.0, 117.1, 112.2, 55.3, 25.1 ppm; HRMS (ESI-TOF) $m/z$: [C$_{43}$H$_{32}$O$_2$]$^+$ calcd. [C$_{43}$H$_{32}$O$_2$] 580.2397; found 580.2389.

**7-methylene-6,8,14-triphenyl-7,14-dihydrobenzo[m]tetraphene-2,12-diol (7b).** A 100 mL Schlenk flask was charged under N$_2$ with **6** (0.375 g, 0.65 mmol) in dry DMF (16 mL). NaSEt (0.540 g, 6.5 mmol) was added under N$_2$ as a solid in one portion. The reaction mixture was stirred under N$_2$ for 3 h at 153 °C. The reaction mixture was quenched with 1M HCl, causing the crude product to precipitate. The crude solid was isolated by filtration and washed with 1M HCl (50 mL) and H$_2$O (100 mL). The crude solid was dissolved in a minimum amount of CH$_2$Cl$_2$ and precipitated by trituration with hexanes yielding **7b** (0.200 g, 0.36 mmol, 56%) as a colorless solid. $^1$H NMR (600 MHz, CD$_2$Cl$_2$) $\delta$ = 7.95 (d, $J$ = 2.0 Hz, 2H), 7.81 (d, $J$ = 8.0 Hz, 2H), 7.67 (s, 2H), 7.46 (m, 4H), 7.40–7.35 (m, 6H), 7.33–7.30 (m, 2H), 7.16 (dd, $J$ = 8.0 Hz, $J$ = 2.0 Hz, 2H), 7.07–7.02 (m, 3H), 6.74 (s, 1H), 5.55 (s, 2H), 4.89 (s, 2H) ppm; $^{13}$C{$^1$H} NMR (151 MHz, CD$_2$Cl$_2$) $\delta$ = 155.0, 143.6, 143.1, 139.8, 136.4, 135.6, 134.7, 132.1, 131.4, 130.6 (2C), 128.8, 128.6, 128.5, 128.5, 127.0, 127.0, 125.8, 118.4, 106.9, 42.9 ppm; (ESI-TOF) $m/z$: [C$_{41}$H$_{27}$O$_2$]$^+$ calcd. [C$_{41}$H$_{27}$O$_2$] 551.2017; found 551.2009.

**7-methylene-6,8,14-triphenyl-7,14-dihydrobenzo[m]tetraphene-2,12-diyl bis(trifluoromethanesulfonate) (8b).** A 100 mL Schlenk flask was charged under N$_2$ with **7b** (0.190 g, 0.34 mmol) in dry CH$_2$Cl$_2$ (34 mL). The reaction mixture was cooled to 0 °C. Et$_3$N (0.425 g, 4.2 mmol) was added dropwise under N$_2$ and the reaction mixture stirred at 0 °C for 15 min. Trifluoromethanesulfonic anhydride (0.593 g, 2.1 mmol) was added dropwise under N$_2$. The reaction mixture was warmed to 24 °C and stirred for 1.5 h at 24 °C. The reaction mixture was concentrated on a rotary evaporator. The crude solid was dissolved in a minimum amount of 1:1 hexanes/CH$_2$Cl$_2$, filtered over a short pad of SiO$_2$, and concentrated on a rotary evaporator yielding **8b** (0.277 g, 0.34 mmol, 99%) as a colorless solid. $^1$H NMR (600 MHz, CD$_2$Cl$_2$) $\delta$ = 8.52 (d, $J$ = 2.0 Hz, 2H), 8.01 (d, $J$ = 8.0 Hz, 2H), 7.84 (s, 2H), 7.49–7.36 (m, 14H), 7.13–7.06 (m, 3H), 6.71 (s, 1H), 5.05 (s, 2H) ppm; $^{13}$C{$^1$H} NMR (151 MHz, CD$_2$Cl$_2$) $\delta$ = 148.8, 142.8, 142.2, 140.2, 138.9, 136.1, 132.3, 132.2, 130.9 (2C), 130.4, 129.3, 128.7, 128.6, 127.8, 127.6, 127.6, 120.6, 120.5, 118.5, 116.2, 43.9 ppm; (ESI-TOF) $m/z$: [C$_{43}$H$_{27}$O$_6$F$_6$S$_2$]$^+$ calcd. [C$_{43}$H$_{27}$O$_6$F$_6$S$_2$] 817.1148; found 817.1152.

**2,2'-(7-methylene-6,8,14-triphenyl-7,14-dihydrobenzo[$m$]-tetra-phene-2,12-diyl)bis(4,4,5,5-tetramethyl-1,3,2-dioxaborolane) (9b).** A 50 mL Schlenk flask was charged under N$_2$ with **8b** (0.130 g, 0.16 mmol), bis(pinacolato)diboron (0.254 g, 0.96 mmol), and KOAc (0.300 g, 2.88 mmol) in dry dioxane (15 mL). The reaction mixture was degassed by sparging with N$_2$ for 20 min before [1,1'-bis(diphenylphosphino)-ferrocene]dichloropalladium(II) (0.013 g, 0.02 mmol) was added under N$_2$. A reflux condenser was attached and the reaction mixture was stirred under N$_2$ for 18 h at 80 °C. The reaction mixture was concentrated on a rotary evaporator. Column chromatography (SiO$_2$; CH$_2$Cl$_2$) yielded **9b** (0.096 g, 0.12 mmol, 78 %) as a colorless solid. $^1$H NMR (600 MHz, CD$_2$Cl$_2$) $\delta$ = 9.24 (s, 2H), 7.89–7.85 (m, 4H), 7.77 (s, 2H), 7.53 (d, $J$ = 8.0 Hz, 4H), 7.43 (t, $J$ = 8.0 Hz, 4H), 7.38–7.34 (m, 4H), 7.28 (s, 1H), 7.08–7.01 (m, 3H), 4.90 (s, 2H), 1.49 (s, 24H) ppm; $^{13}$C{$^1$H} NMR (151 MHz, CD$_2$Cl$_2$) $\delta$ = 144.3, 143.1, 139.8, 139.6, 137.0, 135.0, 134.7, 132.7, 131.1, 130.7, 130.5, 130.2, 129.1, 128.6, 128.5, 128.4, 127.2, 126.8, 125.7, 84.6, 42.7, 25.5, 25.4 ppm; (ESI-TOF) $m/z$: [C$_{53}$H$_{51}$O$_4$B$_2$]$^+$ calcd. [C$_{53}$H$_{51}$O$_4$B$_2$] 773.3968; found 773.3961.

**2,12-dibromo-7-methylene-6,8,14-triphenyl-7,14-dihydrobenzo-[$m$]tetraphene (1b).** A 25 mL sealable Schlenk tube was charged under N$_2$ with **9b** (0.040 g, 0.05 mmol) and CuBr$_2$ (0.070 g, 0.31 mmol) in THF (1 mL), MeOH (2 mL), and H$_2$O (2 mL). The reaction mixture was degassed by sparging with N$_2$ for 20 min. The reaction mixture was sealed under N$_2$ and stirred for 18 h at 120 °C. The reaction mixture was concentrated on a rotary evaporator, diluted with H$_2$O (10 mL), and extracted with CH$_2$Cl$_2$ (30 mL). The combined organic phases were washed with H$_2$O (10 mL), saturated aqueous NaCl (10 mL), dried over MgSO$_4$, and concentrated on a rotary evaporator. Column chromatography (SiO$_2$; 4:1 hexane/CH$_2$Cl$_2$) yielded **1b** (0.034 g, 0.05 mmol, 96 %) as a colorless solid. $^1$H NMR (600 MHz, CD$_2$Cl$_2$) $\delta$ = 8.78 (s, 2H), 7.79–7.73 (m, 4H), 7.63 (d, $J$ = 8.0 Hz, 2H), 7.47–7.26 (m, 12H), 7.11–7.05 (m, 3H), 6.82 (s, 1H), 4.93 (s, 2H) ppm; $^{13}$C{$^1$H} NMR (151 MHz, CD$_2$Cl$_2$) $\delta$ = 142.9, 142.6, 139.2, 139.1, 135.9,

135.1, 132.0, 131.7, 131.1, 130.8, 130.4, 130.1, 129.0, 128.6, 128.5, 127.4, 127.2, 126.8, 126.7, 121.7, 42.8 ppm; HRMS (EI-TOF) m/z: $[C_{41}H_{26}Br_2]^+$ calcd. $[C_{41}H_{26}Br_2]$ 678.0381; found 678.0381.

**5-oGNR Growth on Au(111) Surfaces.** 5-oGNRs were grown on Au(111)/mica films under UHV conditions. Atomically clean Au(111) surfaces were prepared through iterative Ar$^+$ sputter/anneal cycles. Sub-monolayer coverage of **1b** on atomically clean Au(111) was obtained by sublimation at crucible temperatures of 453-473 K using a Knudsen cell evaporator. After deposition the surface temperature was slowly ramped ($\leq 2$ K min$^{-1}$) to 453 K and held at this temperature for 15 min to induce the radical-step growth polymerization, then ramped slowly ($\leq 2$ K min$^{-1}$) to 623 K and held there for 15 min to induce cyclodehydrogenation.

**Scanning Tunnelling Microscopy and Spectroscopy.** All STM experiments were performed using a commercial OMICRON LT-STM operating at $T = 4$ K using PtIr STM tips. STM tips were optimized for scanning tunnelling spectroscopy using an automated tip conditioning program.[42] d$I$/d$V$ measurements were recorded with CO-functionalized STM tips using a lock-in amplifier with a modulation frequency of $f = 455$ Hz and a modulation amplitude of $V_{ac} = 10$ mV. d$I$/d$V$ point spectra were recorded under open feedback loop conditions. d$I$/d$V$ maps were collected under constant height conditions. Peak positions in d$I$/d$V$ point spectroscopy were determined by fitting the spectra with Lorentzian peaks. Each peak position is based on an average of ~10 spectra collected on various GNRs with different tips, all of which were first calibrated to the Au(111) Shockley surface state.

**Calculations.** First-principles DFT calculations in the LDA and LSDA approximations were implemented using the Quantum Espresso package.[43] We used Norm-conserving (NC) pseudopotentials with a 60 Ry energy cut-off and 0.005 Ry Gaussian broadening. To ensure the accuracy of our results, a sufficiently large vacuum region was included in the supercell calculation. All of the dangling bonds at the edge of the carbon skeleton were hydrogenated. The structures were first fully relaxed until all components of the force were smaller than 0.01 eV/ Å.

## ASSOCIATED CONTENT

### Supporting Information

The Supporting Information is available free of charge on the ACS Publications website.

Characterization of **1b**
X-ray crystal structure data for **1b** [CCDC 2130874]
X-ray crystal structure data for **8b** [CCDC 2130875]

### Accession Codes

CCDC 2130874 and CCDC 2130875 contain the supplementary crystallographic data **1b** and **8b**, respectively. These data can be obtained free of charge via www.ccdc.cam.ac.uk/data_request/cif, or by emailing data_request@ccdc.cam.ac.uk, or by contacting The Cambridge Crystallographic Data Centre, 12 Union Road, Cambridge CB2 1EZ, UK; fax: +44 1223 336033.

## AUTHOR INFORMATION

### Corresponding Author

* Felix R. Fischer; Email: ffischer@berkeley.edu
* Steven G. Louie; Email: sglouie@berkeley.edu

### Author Contributions

All authors have given approval to the final version of the manuscript.
‡These authors contributed equally.

### Notes

The authors declare no competing financial interest.

## ACKNOWLEDGMENT

This work was primarily funded by the US Department of Energy (DOE), Office of Science, Basic Energy Sciences (BES), Materials Sciences and Engineering Division under contract DE-AC02-05-CH11231 (Nanomachine program KC1203) (molecular design, tight-binding studies, topological states analysis) and contract DE-SC0023105 (surface growth). Research was also supported by the Office of Naval Research under award N00014-19-1-2503 (STM characterization), the National Science Foundation under grant nos. CHE-2203911 (STS analysis) and DMR-1926004 (DFT calculations). Part of this research program was generously supported by the Heising-Simons Faculty Fellows Program at UC Berkeley. STM instruments are supported in part by the Office of Naval Research under award N00014-20-1-2824. This research used resources of the National Energy Research Scientific Computing Center (NERSC), a U.S. Department of Energy Office of Science User Facility operated under Contract No. DE-AC02-05CH11231. Computational resources were also provided by the NSF TACC Frontera and NSF through ACCESS resources at the NICS (stampede2). E.C.H.W. acknowledges support from the Croucher Foundation through a Croucher Scholarship for Doctoral Study. We thank Dr. Hasan Çelik and the UC Berkeley NMR facility in the College of Chemistry (CoC-NMR) for assistance with spectroscopic characterization. Instruments in the CoC-NMR are supported in part by National Institutes of Health (NIH) award no. S10OD024998.